\documentclass[aps,pra,twocolumn,showpacs,floatfix]{revtex4}
\usepackage{graphicx}
\begin{document}

\title{Universal and nonuniversal contributions to block-block entanglement 
in many-fermion systems}

\author{V. V. Fran\c{c}a}
\author{K. Capelle}
\email{capelle@if.sc.usp.br}
\affiliation{Departamento de F\'{\i}sica e Inform\'atica,
Instituto de F\'{\i}sica de S\~ao Carlos,
Universidade de S\~ao Paulo,
Caixa Postal 369, 13560-970 S\~ao Carlos, SP, Brazil}
\date{\today}

\begin{abstract}
We calculate the entanglement entropy of blocks of size $x$ embedded
in a larger system of size $L$, by means of a combination of analytical and 
numerical techniques. The complete entanglement entropy in this case is a sum 
of three terms. One is a universal $x$ and $L$-dependent term, first predicted 
by Calabrese and Cardy, the second is a nonuniversal term arising from the
thermodynamic limit, and the third is a finite size correction. We give an
explicit expression for the second, nonuniversal, term for the one-dimensional 
Hubbard model, and numerically assess the importance of all three contributions
by comparing to the entropy obtained from fully numerical diagonalization 
of the many-body Hamiltonian. We find that finite-size corrections are very
small. The universal Calabrese-Cardy term is equally small for small blocks,
but becomes larger for $x>1$. In all investigated situations, however, the
by far dominating contribution is the nonuniversal term steming from the 
thermodynamic limit.
\end{abstract}

% 03.67.Mn Entanglement production, characterization, and manipulation
% 71.10.Fd Lattice fermion models (Hubbard model, etc.)
% 03.65.Ud Entanglement and quantum nonlocality
% 71.10.Pm Fermions in reduced dimensions

\maketitle

\newcommand{\be}{\begin{equation}}
\newcommand{\ee}{\end{equation}}
\newcommand{\bea}{\begin{eqnarray}}
\newcommand{\eea}{\end{eqnarray}}
\newcommand{\bi}{\bibitem}

\renewcommand{\r}{({\bf r})}
\newcommand{\rp}{({\bf r'})}

\newcommand{\ua}{\uparrow}
\newcommand{\da}{\downarrow}
\newcommand{\la}{\langle}
\newcommand{\ra}{\rangle}
\newcommand{\dg}{\dagger}

\section{Introduction}
\label{intro}

Entanglement is one of the most studied and least intuitive features of
quantum mechanics. Many aspects of it are still not fully understood.
Part of the difficulty is that in itself entanglement 
is not an observable quantity. Rather, entanglement is a property of the 
quantum mechanical state, defined with respect to some set of degrees of 
freedom. For different degrees of freedom, and different states, entanglement
is characterized and quantified in different ways.

For mixed states, described by a density operator, many alternative measures
of entanglement are still under study. For pure states in a bipartite system, 
described by a 
wave function, on the other hand, a near-consensus has emerged that the
entanglement entropy is a suitable entanglement measure. Having identified 
a suitable measure, the task at hand then changes to evaluating it and 
analysing its behaviour in various physical systems, in order to extract 
information that can be useful in quantum information processsing and 
computing.

The present paper is concerned with this task in the particular case of
strongly interacting electrons in a finite-size chain. Our interest is in
separating universal and system-specific contributions to the entanglement
entropy, quantifying their relative importance, and investigating their
behaviour as a function of system parameters. Specifically, we consider 
a quantum chain of length $L$ divided in a subsystem $A$ of size 
$x$, and a subsystem $B$ of size $L-x$, and calculate the entanglement 
entropy \cite{bennett}
\be
S(x,L)=-Tr[\rho_A \log_2 (\rho_A)],
\ee
where the reduced density matrix $\rho_A=Tr_B [\rho]$ is obtained from the 
density matrix of the full system, $\rho$, by tracing over the degrees of 
freedom of subsystem $B$. 
For interacting many-particle systems the full density matrix is almost 
impossibly difficult to obtain. For suitable model Hamiltonians, and not
too many particles, however, fully numerical diagonalization is within
reach, and can be used to calculate $\rho_A$ and $S$. Before embarking on
such a numerical calculation for a specific system, however, it is useful 
to recall general properties of $S(x,L)$ that were uncovered in ground-breaking
analytical work of Calabrese and Cardy \cite{cc}.

These authors find that the entanglement entropy of a subsystem of size $x$, 
embedded in a larger gapless system of size $L\gg x$, consists of two distinct 
terms: a universal term depending only on $x$ and $L$, and a nonuniversal term 
that depends on system-specific parameters, but is independent of $x$ and 
$L$ \cite{cc,korepin}. Analytical expressions for the universal 
term were obtained by Calabrese and Cardy (CC) \cite{cc}, found to be
in agreement with partial results obtained earlier in Refs.~\cite{korepin,affleck,holzhey}, and were further analysed in, {\em e.g.}, Refs.~\cite{sarandyCC,laflorencie1,laflorencie2,schollwoeck}. 

For periodic boundary conditions and $L \gg x \gg 1$, these authors find
\be
S(x,L)= \frac{c}{3}
\log_2\left[\frac{L}{\pi}\sin\left(\frac{\pi x}{L}\right)\right]+s_1,
\label{cc1}
\ee
where $c$ is the central charge (conformal anomaly) of the system and $s_1$ 
is a nonuniversal term whose magnitude and dependence on system parameters 
remain undetermined in the approach of Refs.~\cite{cc,korepin,affleck,holzhey}.
If the condition $L \gg x \gg 1$ is not satisfied there may be additional
finite-size corrections, not contained in the CC analysis.
While the identification of universal terms is one of the principle 
goals of statistical physics, any quantitative application to realistic 
models or to actual materials and devices, depends crucially on information 
about the nonuniversal terms. With a view on future realizations of quantum 
computing and quantum information processing in systems of interacting 
particles, we therefore now embark on the task to extract information about 
the nonuniversal function $s_1$ and on possible finite-size corrections 
for realistic models of such systems.

In this paper, we focus on the one-dimensional fermionic Hubbard model.
For this model, we (i) numerically assess the magnitude 
of the universal Cardy-Calabrese (CC) term and the nonuniversal $s_1$ term for 
realistic values of system parameters, arriving at the unexpected conclusion 
that the universal 
term is only a small correction to the much larger nonuniversal term; (ii)
obtain an analytical expression for $s_1$ of the Hubbard model, allowing us 
to study its dependence on various system parameters; and (iii) compare the 
analytical expression with numerical results obtained by full diagonalization 
of the many-body Hamiltonian, finding satisfactory agreement, both for 
single-site and block-block entanglement, in interacting and noninteracting 
systems of various sizes and densities.

\section{Universal contribution to the entropy versus exact entropy}
\label{univentrop}

In this section we numerically calculate the exact entanglement entropy of the 
one-dimensional finite-size Hubbard model and compare it to the analytical
prediction made by keeping only the universal term in the CC formula,
\be
S^{\rm univ}(x,L)=\frac{c}{3}\log_2\left[\frac{L}{\pi}\sin\left(\frac{\pi x}{L}\right)\right].
\label{univent}
\ee

The one-dimensional Hubbard model, one of the most widely used models of 
strongly interacting particles \cite{essler,takahashi,zvegbyn}, is described 
by the Hamiltonian
\be
\hat{H}=-t\sum_{i,\sigma}(\hat{c}_{i\sigma}^\dagger\hat{c}_{i+1,\sigma}+H.c.)+U\sum_i\hat{c}_{i\uparrow}^\dagger\hat{c}_{i\uparrow}\hat{c}_{i\downarrow}^\dagger\hat{c}_{i\downarrow},
\ee
where $t$ is the hopping between neighbouring sites, $U$ is the on-site
particle-particle interaction and $\hat{c}_{i\sigma}^\dagger$ and 
$\hat{c}_{i\sigma}$ are (fermionic) creation and annihilation operators of 
particles at site $i$ with spin $\sigma$. The system described by this 
Hamiltonian is completely characterized by its size $L$, interaction $U$ 
\cite{footnote1} and either the particle number $N\leq 2L$ or the particle 
density $n=N/L$ \cite{footnote1b}. For small $L$, this Hamiltonian
can be diagonalized numerically. Since this involves no approximation other
than the use of finite-precision numbers on a computer, we follow common
terminology and denote this as exact diagonalization. The resulting 
eigenfunctions can be used to construct the density matrix, and from this 
the entanglement entropy can be extracted. 

A quantitative comparison between the resulting $S^{\rm exact}$ and 
$S^{\rm univ}$, as given by Eq.~(\ref{univent}), is presented in 
Table~\ref{table1}, for various different choices of system parameters 
and $x=1$. Interestingly, the universal term makes only 
a very small, negative, contribution to the exact single-site entropy.

\begin{table}
\caption{Universal contribution to the single-site ($x=1$) entanglement 
of a chain of size $L$ with periodic boundary conditions, compared to data 
from numerically exact full diagonalization, at $U=4$. The last column is the
deviation, in percent, of the universal contribution from the
exact numerical value.}
\label{table1}
\begin{ruledtabular}
\begin{tabular}{ccccc}
$n$ &L  &$S^{\rm univ}$ & $S^{\rm exact}$ & deviation (\%) \\
\hline
&4  & -0.101&1.541&106.5\\
$0.5$&8 &  -0.025&1.564&101.6\\
&12 &  -0.011&1.576&100.7\\
\hline
&4  & -0.050&1.594&103.1\\
$1.0$&8 &  -0.012&1.701&100.7\\
&12 &  -0.005&1.718&100.3\\
\end{tabular}
\end{ruledtabular}
\end{table}

We conclude from this analysis that for the single-site entanglement the 
universal CC term is hardly relevant quantitatively: when $L$ is large 
enough for the full CC formula to become asymptotically exact (recall that 
it was derived for $L\gg x$), the universal term is already a vanishingly 
small correction to the nonuniversal term. 

From Eq.~(\ref{univent}) it is clear that the universal term increases 
as a function of block size $x$. The interesting question is then by how 
much it grows relative to the nonuniversal, system-specific, term $s_1$. 
In order to investigate this quantitatively, we fix $L$ at 10 sites, and 
calculate $S(x,L=10)$ as a function of the size $x$ of the subsystem $B$. 
Benchmark data for comparison at $x>1$ are extracted from Ref.~\cite{deng}, 
which deals with block-block entanglement in the
extended Hubbard model. By setting the parameter $V$ of that model equal
to zero, reference values for $S(x,L=10)$ can be extracted from
Figure 3 of that work. Below we do refer to these data as ``exact"
because they were also obtained by numerically exact full diagonalization,
but we note that we extracted them graphically from Figure 3 of 
Ref.~\cite{deng}. The difference between these ``exact" values and 
$S^{\rm univ}$, which is the effect we are after, is clearly much larger 
than any possible error of the benchmark data, which can therefore safely 
be used for comparison.

\begin{table}[t]
\caption{Universal contribution to the block-block entanglement of a chain 
of size $L=10$ with $U=0$ and $n=1$, compared to values for the full entropy
extracted for the same system from Ref.~\cite{deng}. The last column is the
deviation, in percent, of the universal contribution from the 
reference value.}
\label{table2}
\begin{ruledtabular}
\begin{tabular}{cccc}
x & $S^{\rm univ}$ & $S^{exact}$ & deviation (\%) \\
\hline
1 & -0.016 &  2.00&100.8\\
2 &0.602 &   2.69&77.6\\
3 & 0.910 &  3.02& 69.9\\
4 &1.065 &  3.17 &66.4\\
5 & 1.114 &   3.22&65.4\\
\end{tabular}
\end{ruledtabular}
\end{table}

The data in Table~\ref{table2} show that for larger blocks the universal term
makes a more noticeable contribution to the block-block entanglement entropy.
While for $x=1$ the deviation of the universal term from the numerical value 
is very similar to that of Table~\ref{table1}, it becomes smaller for $x>1$. 
However, quantitatively, it is still amounts to less than half of the exact 
entropy. As an example, even for $x=5$, {\em i.e.}, a maximal subsystem half 
the size of the complete system, it is still only about one third of 
$S^{exact}$. Part of this difference must be due to $s_1$.

\begin{figure}[t] 
\centering
\includegraphics[height=7cm]{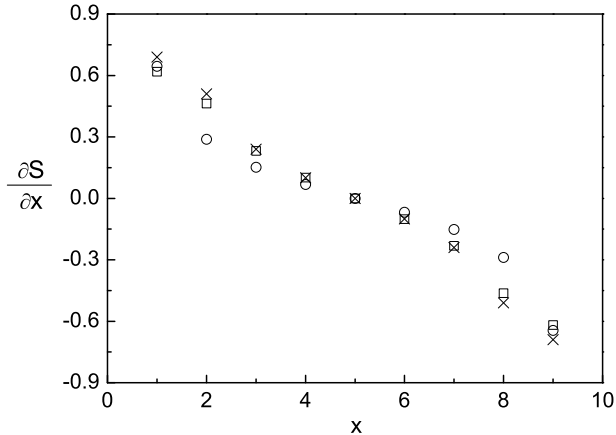}
\caption{Derivative of the block-block entanglement entropy with respect 
to block size, obtained analytically from Eq.~(\ref{cc1}) [open circles],
numerically from five data points obtained from Eq.~(\ref{cc1}) [open squares],
and numerically from the five data points in Table~\ref{table2}.}
\label{fig1}
\end{figure}

The $s_1$ term can be eliminated by studying not $S$, but 
its derivative as a function of block size $x$. This behaviour can be 
evaluated in various ways. First, we calculate analytically the derivative 
$\partial S(x,L)/\partial x$ of the CC expression (\ref{cc1}), by treating
$x$ as a continuous variable. Results for integer values of $x$ are represented
by open circles in Fig.~\ref{fig1}. Note that since $s_1$ is taken to be a 
constant, this derivative samples only the universal term. Second, we calculate
the numerical derivative of the five data points for $x=1,..,5$ collected in 
Table~\ref{table2}. Results are represented by crosses in Fig.~\ref{fig1}.
A direct comparison between both sets of data is marred by the intrinsic 
inaccuracy of a numerical derivative. To minimize this problem, we also 
obtained the derivative of the CC formula (\ref{cc1}) by evaluating that
expression numerically at $x=1,..,5$ and taking the numerical derivative of
the resulting values. This set of data, represented by open squares, is 
directly comparable to the numerical derivative of the benchmark data in
Table~\ref{table2}.

The deviation between the open circles and open squares is thus between
analytical and numerical 5-point derivatives of the same function. The
difference between the open squares and the crosses is between the CC 
prediction of the trend as a function of $x$ and the numerical results,
both obtained from five data points. 
Evidently, the behaviour of the CC expression and of the numerical data is
very similar. The small differences remaining between crosses and squares
are due to finite-size corrections to the CC expression (more on these
below) and the intrinsic error bar of the reference data from 
Table~\ref{table2}.

Since the behaviour of the derivative $\partial S/\partial x$ is very closely 
reproduced by the CC expression, we conclude that the much larger differences 
observed in Table~\ref{table2} for the entropy itself, must be almost entirely 
due to the term $s_1$. This analysis thus points to the importance of 
nonuniversal terms, which remain undetermined in the CC approach. 

Additionally, it should be
noted that the CC formula was derived for large $L\gg x$, while in order
to be able to compare to data from full numerical diagonalization we 
evaluate it for $L<13$. Finite-size corrections to the CC formula are
another possible explanation for the large fraction of the exact entropy 
not recovered by the universal term only. In the next sections we attempt
to disentangle and quantify these two distinct effects, by deriving 
an analytical expression for $s_1$ of the Hubbard model, and comparing it
to the same set of exact data.

\section{Analytical expression for the block-entanglement entropy}
\label{analytical}

In the thermodynamic limit, $L\to\infty$, the CC expression (\ref{cc1}) 
reduces to
\be
S(x,L\to \infty)= \frac{c}{3}\log_2(x)+s_1,
\label{ccL}
\ee
from which the function $s_1$ of the model under study can be
determined once the entanglement entropy of this model is known in the
thermodynamic limit. Since $s_1$ does not depend on $x$, we are
free to evaluate it for any convenient value of $x$. In the special case
of single-site entanglement ($x=1$), the logarithm on the right-hand side
vanishes, and
\be
s_1 = S(x=1,L\to \infty),
\label{tdlim}
\ee
which implies
\be
\frac{S(x,L\to \infty)}{S(1,L\to \infty)}=1 + \frac{c}{3}\frac{\log_2(x)}{s_1},
\ee
This identification neglects possible finite-size corrections arising from the 
fact that the CC expression was derived only for $L\gg x\gg1$. The difference
between numerical data and predictions of the preceding equation allows one to 
estimate the size of such corrections.

Next, we apply this procedure to the Hubbard chain. Recent research has 
resulted in a complete physical picture of and explicit expressions for 
$S(x=1,L\to \infty)$ \cite{gu,larsson,franca1}.
Specifically, the single-site entanglement entropy for the Hubbard model in
the absence of external electric or magnetic fields is given by
\cite{gu,larsson,franca1}
\bea
S(x=1,L\to \infty;n,U)=-2\left(\frac{n}{2}-\frac{\partial e}{\partial U}\right)\log_2\left[\frac{n}{2}-\frac{\partial e}{\partial U}\right]\label{hom.hu}
\nonumber \\
-\left(1-n+\frac{\partial e}{\partial U}\right)\log_2\left[1-n+\frac{\partial e}{\partial U}\right],
\label{homent}
\eea
where the ground-state energy per site, $e=E_0(n,U)/L$, can be obtained from
the Bethe-Ansatz integral equations \cite{essler,takahashi,zvegbyn,liebwu}.

By combining this Bethe-Ansatz based expression for $S(x=1,L\to \infty)$
with the CC formula, based on conformal field theory, we obtain
\bea
S(x,L;n,U)=
\frac{c}{3}\log_2\left[\frac{L}{\pi}\sin\left(\frac{\pi x}{L}\right)\right]
\nonumber \\ 
-2\left(\frac{n}{2}-\frac{\partial e}{\partial U}\right)\log_2\left[\frac{n}{2}-\frac{\partial e}{\partial U}\right]
\nonumber \\ 
-\left(1-n+\frac{\partial e}{\partial U}\right)\log_2\left[1-n+\frac{\partial e}{\partial U}\right],
\label{shu}
\eea
which is the sum of a universal term, depending only on
geometry ($x$ and $L$), and a term depending on specific system
parameters ($n$ and $U$). This explicit expression allows us to investigate
the actual size of each of these terms under realistic circumstances, by
obtaining $e(n,U)$ numerically from the Bethe-Ansatz integral equations, 
and evaluating Eq.~(\ref{shu}) as a function of $x$, $L$, $n$ and $U$.

\section{Full entropy versus exact entropy}
\label{fullentrop}

\begin{figure}[t]
\centering
\includegraphics[height=7cm]{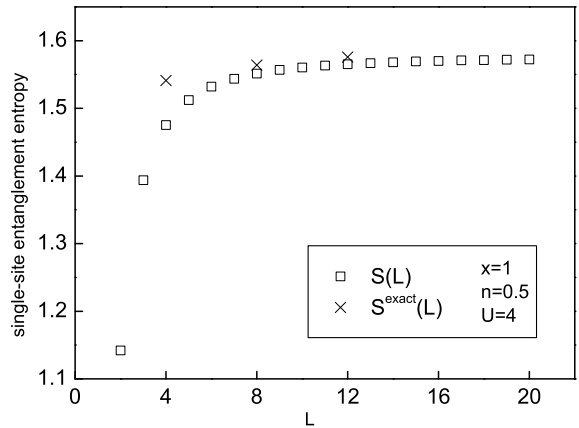}
\caption{Single-site entanglement entropy as a function of system size $L$
for the Hubbard model. Open squares: analytical results from our
Eq.~(\ref{shu}). Crosses: numerical data obtained by diagonalizing 
the many-body Hamiltonian.}
\label{fig2}
\end{figure}

Figure~\ref{fig2} contains a comparison of our numerical data for $x=1$
with our analytical expression (\ref{shu}) for the specific case of a Hubbard
chain with $U=4$ and $n=0.5$. Exact data are given only for
$L=4,8,12$, because odd particle numbers $N$ would result in a finite
magnetization (which is not included in Eq.~(\ref{homent})) and $L\geq 13$ is
already too large for full exact diagonalization on our computing equipment.
The available data, however, are clearly sufficient to conclude that trend 
and magnitude are the same for both analytical and numerical data.

The quantitative deviation observed in Fig.~\ref{fig2} between analytical and 
numerical results for small $L$ is due to the fact that the CC formula was 
derived for large $L$, whereas data from full numerical diagonalization 
are only available for $L<13$. For small $L$, there may be additional 
$L$-dependent terms in the full expression for the entropy, which go to zero 
as $L$ increases. These are the finite-size effects, referred to above. The 
data in Fig.~\ref{fig2} show that already for $L$ as small as 12, 
such possible small-$L$ corrections are negligible. 

\begin{figure}[t]
\centering
\includegraphics[height=7cm]{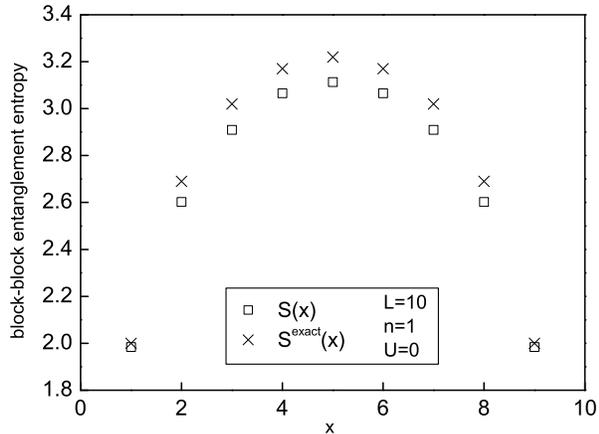}
\caption{Block-block entanglement entropy as a function of the block size $x$ 
for the noninteracting Hubbard model ($L=10$, $U=0$ and $n=1$). For $x>L/2$ 
we obtained $S$ from the symmetry relation $S(x,L)=S(L-x,L)$, and for
$x<L/2$ is calculated from Eq.~(\ref{shu}). Benchmark data have been
extracted from Ref.~\cite{deng}. %The table in the inset compares 
%data for noninteracting systems ($U=0$) to data for repulsively ($U=2$) 
%and attractively ($U=-2$) interacting systems with block size $x=4$.
}
\label{fig3}
\end{figure}

Figure~\ref{fig3} extends this analysis to larger block sizes, by evaluating
Eq.~(\ref{shu}) as a function of $x$, and comparing to the same set of 
reference data at $x>1$ and $U=0$, used in Sec.~\ref{univentrop}. 
%The inset 
%extends this comparison to available data \cite{deng} for repulsively ($U>0$) 
%and attractively ($U<0$) interacting systems. 
The overall agreement between 
expression (\ref{shu}) and the benchmark data is rather satisfactory.
The remaining deviations now have two distinct sources. 
One is, as above, the use of the CC expression at rather small $L$. The
other is the imprecision in the extraction of the benchmark data from the
figure presented in Ref.~\cite{deng}. Still, there can hardly be any doubts 
that the dependence on block size $x$ is reproduced correctly. Clearly, in 
Eq.~(\ref{shu}) this dependence comes exclusively from the universal term, 
which makes a much more pronounced contribution for $x>1$ than it made for 
$x=1$. 

\begin{table}[t]
\caption{Same as Table~\ref{table1}, but with $S^{\rm univ}(x,L)$ replaced
by our expression (\ref{shu}), comprising $S^{\rm univ}(x,L)$ and expression
(\ref{homent}) for $s_1(n,U)$.}
\label{table3}
\begin{ruledtabular}
\begin{tabular}{ccccc}
 $n$ & $L$  & $S(x=1,L)$ & exact & deviation (\%) \\
\hline
& 4  & 1.475&1.541&4.3\\
$0.5$ & 8 & 1.551&1.564&0.8\\
& 12 &  1.565&1.576&0.7\\
\hline
& 4  & 1.678&1.594&5.3\\
$1.0$ &8 & 1.716&1.701&0.9\\
&12 &  1.723&1.718&0.3\\
\end{tabular}
\end{ruledtabular}
\end{table}

\begin{table}[t]
\caption{Same as Table~\ref{table2}, but with $S^{\rm univ}(x,L)$ replaced
by our expression (\ref{shu}), comprising $S^{\rm univ}(x,L)$ and expression
(\ref{homent}) for $s_1(n,U)$.}
\label{table4}
\begin{ruledtabular}
\begin{tabular}{cccc}
x &   $S(x,L)$ &exact & deviation (\%) \\
\hline
1 & 1.984  & 2.00&0.8\\
2 & 2.602  & 2.69& 3.3\\
3 & 2.910  & 3.02& 3.6\\
4 &3.065  & 3.17 &3.3\\
5 & 3.114  & 3.22& 3.3\\
\end{tabular}
\end{ruledtabular}
\end{table}

Tables~\ref{table3} and \ref{table4} compare expression (\ref{shu}) to
benchmark data at $x=1$ and $x>1$. The last column of Tables~\ref{table3} 
and \ref{table4} shows that expression (\ref{shu}) practically exhausts the 
exact entropy, both for single-site entanglement and for block-block 
entanglement. As before,
we attribute the remaining small differences, of order $\sim 1\%$, to
finite-size corrections, contained in the numerical data for small $L$ but
not in the CC expression derived for large $L$. Since for some values of
$n$ the full expression (\ref{shu}) predicts more than $100\%$ of the exact 
entropy, these finite-size corrections must alternate their sign as a function 
of $n$.

As a second test, we have also fitted the exact data for $L=10$ and $x\leq 5$
in Tables \ref{table2} and \ref{table4} with expression (\ref{cc1}), treating 
$c$ and $s_1$ as fitting parameters. The result is $c = 2.17\pm0.02$ and 
$s_1 = 2.02\pm0.007$. Since in the situation of these Tables the exact values 
are known to be $s_1=c=2$, this fit again illustrates the smallness of 
finite-size corrections to the CC formula (\ref{cc1}) for $x$ and $L$ 
outside the range $L\gg x\gg1$, where it becomes exact.

\section{Conclusions and Outlook}
\label{conclusions}

We can summarize our combined analytical-numerical analysis in the statement 
that the block-block entanglement entropy for {\em any} system size $L$ and 
block size $x$ is given by
\bea
S(x,L;n,U) &=& S^{\rm univ}(x,L) \nonumber\\
 + S(x=1,L\to \infty;n,U) &+& \Delta S(x,L;n,U),
\label{stotal}
\eea
where the first term is the universal CC term of Eq.~(\ref{univent}), 
depending on $x$ and $L$ only, and the second is the system-specific 
term $s_1$, depending on $n$ and $U$, which we extract from the thermodynamic 
limit (\ref{tdlim}). The third by definition comprises all possible finite-size
corrections not resulting from the CC analysis and neither contained in the 
infinite-size limit leading to the identification 
$s_1(n,U)=S(x=1,L\to \infty;n,U)$. 

By comparing this full expression to the data in Tables~\ref{table1} to 
\ref{table4}, and the trends visible in Figs.~\ref{fig1} to \ref{fig3},
we conclude that the nonuniversal term $s_1=S(x=1,L\to \infty;n,U)$ 
quantitatively dominates the physics of the entanglement entropy of the 
Hubbard model in all investigated situations. For $x=1$, the universal term 
and the finite-size corrections are of comparable magnitude, $O(1 \%)$,
and essentially negligible relative to the nonuniversal term, even for rather 
small $L$. For $x>1$, the universal term is of order $O(10\%)$ of the
full entropy, while finite-size corrections remain $ O(1 \%)$.
The trend as a function of $x$, on the other hand, is not at all affected
by $s_1$, but dominated by the universal term, receiving only small
($\sim 1-10\%$) finite-size corrections from $\Delta S(x,L;n,U)$.

While all of this highlights the intellectual achievement of CC in having
identified the universal contribution to the entanglement entropy, it also
shows that if one wants to quantify the entropy in an actual material or 
device --- a need that arises as soon as one considers using entanglement
as a resource for quantum computing in {\em real} systems --- a detailed
description of system-specific features is unavoidable.

One way to obtain an approximate system-specific description also in systems 
with inequivalent sites is the local-density approximation (LDA) to 
density-functional theory (DFT), which locally applies results obtained in a 
spatially homogeneous system (with constant density $n$) in order to simulate 
the corresponding inhomogeneous system (with spatially varying density $n_i$).
This approximation is commonly applied in {\em ab initio} calculations
of the electronic structure of atoms, molecules and solids
\cite{dftbook,parryang,kohnrmp,perdewoverview}, where the local approximation
(or one of its refinements) is made for the exchange-correlation energy.
In earlier work we proposed to apply the same strategy also to
the Hubbard model \cite{balda}, and suggested a simple local-density 
approximation for the single-site entanglement entropy \cite{franca2}.
(For related applications of DFT concepts to the study of entanglement,
see Refs.~\cite{sarandy,damico}.)

That approximation was constructed specifically for $x=1$. The present 
equation (\ref{stotal}), valid for all $x$, suggests two ways to extend the
validity of the local-density approximation to block-block entanglement.
One is to simply add to the LDA of Ref.~\cite{franca2} the term 
$M S^{\rm univ}(x,L)$, which the original entropy LDA did not contain. Here
$M=L/x$ is the number of blocks in the system.
The other is to take Eq.~(\ref{stotal}) as entropy of the homogeneous 
reference system on which the local approximation is based. This leads to
\bea
S^{LDA}[x,L;n_i,U]=M S^{\rm univ}(x,L\to\infty)+
\nonumber \\
{1 \over x} S^{LDA}[x=1,L;n_i,U],
\eea
which differs from the {\em ad hoc} correction in the $L$-dependence of the 
first term and the $x$-dependence of the second. Future work directed at 
block-block entanglement in spatially inhomogeneous systems is expected to
identify which of these two extensions is more reliable.
Interesting inhomogeneities, in this context, include impurities, defects,
spatial modulations of the system parameters, confining potentials, etc.

A combination of the methodologies employed in Ref.~\cite{franca2}
and in the present paper thus allows to analyse and quantify the entanglement 
entropy in a wide variety of spatially inhomogeneous many-body systems. 
Such analysis, and an extension of these investigations to spin-polarized
systems, is subject of future work.

{\bf Acknowledgments}
This work was sup\-por\-ted by FAPESP, CNPq and CAPES. We thank Vivaldo
L. Campo Jr. for providing us with his efficient Lanczos subroutine.
%, and Pasquale Calabrese and Marcelo S. Sarandy for useful comments 
% on an earlier version of the manuscript.

%)}]
\end{document}